\documentclass[conference]{IEEEtran}
\IEEEoverridecommandlockouts
\usepackage{cite}
\usepackage{amsmath,amssymb,amsfonts}
\usepackage{algorithm, algpseudocode}
\usepackage{graphicx}
\usepackage{textcomp}
\usepackage{xcolor}
\usepackage[linguistics]{forest}
\usepackage{adjustbox}
\usepackage{soul}
\usepackage{caption}
\usepackage{subcaption}
\usepackage{tikz}
\usepackage{pgfplots}
\pgfplotsset{compat=1.18}
\usepackage{comment}
\usepackage{booktabs}
\usepackage{hyperref}

\begin{document}

\title{Data Certification Strategies for\\ Blockchain-based Traceability Systems
\thanks{This work was partially supported by project SERICS (PE00000014) under the MUR National Recovery and Resilience Plan funded by the European Union - NextGenerationEU, by project FOLOU funded by the European Union’s Horizon Europe research and innovation program (GA: 101084106), and by MIMIT, under FSC project ``Pesaro CTE SQUARE'', CUP D74J22000930008.}
}

 \author{
  \IEEEauthorblockN{Giacomo Zonneveld, Giulia Rafaiani, Massimo Battaglioni and Marco Baldi}
  \IEEEauthorblockA{ Dipartimento di Ingegneria dell’Informazione \\
                    Università Politecnica delle Marche \\
                    Ancona, Italy \\
                    Email: \{g.zonneveld, g.rafaiani, m.battaglioni, m.baldi\}@univpm.it}
}

\maketitle

\begin{abstract}
The use of blockchains for data certification and traceability is now well established in both the literature and practical applications. However, while blockchain-based certification of individual data is clear and straightforward, the use of blockchain to certify large amounts of data produced on a nearly continuous basis still poses some challenges.
In such a case, in fact, it is first necessary to collect the data in an off-chain buffer, and then to organize it, e.g., via Merkle trees, in order to keep the size and quantity of certification data to be written to the blockchain small.
In this paper, we consider a typical system for blockchain-based traceability of a production process, and propose and comparatively analyze some strategies for certifying the data of such a process on blockchain, while maintaining the possibility of verifying their certification in a decentralized way.
\end{abstract}


\section{Introduction}

Blockchain technology, introduced in 2009 to support the first cryptocurrency \cite{nakamoto}, and then generalized to the concept of distributed ledger technology (DLT), now lends itself to many applications other than just monetary transactions.
These include data notarization and certification applications, in which the features of the blockchain are exploited to make it play the role of a notary and prove the existence of a certain amount of data at a certain instant in time (within some temporal resolution).
Typical applications of this use of blockchain are those related to agri-food supply chain traceability \cite{Wang2019,Compagnucci2020} and production processes traceability in general, as well as traceability and data certification for Internet of Things \cite{Fernandez2018} and Industry 4.0 \cite{Bodkhe2020} applications.
Another example of traceability for which blockchain-based solutions are suitable is that of Proof of Attendance, in which the aim is to certify users' participation in various kinds of events, such as cultural events, tourism initiatives, or their attendance to class or work \cite{ardina,dreyfus}.
In such a case, instead of production or supply chain data, the data to be certified will be attendance certificates or similar data.
It should be noted that, in all these cases, for reasons of confidentiality as well as data size, it is not possible to write the data itself onto the blockchain, but only certification data derived from it, from which the data itself cannot be reconstructed.

In fact, blockchain technology provides wide support for data certification.
Thanks to its properties such as immutability, once data are recorded into the blockchain, they become unalterable, which ensures data integrity over time. The transparent and distributed nature of blockchain ensures trust, since anyone can directly verify the information contained in the public ledger.
Given its nature, however, blockchain technology cannot be used as a database. In fact, blockchain is not meant for storing big amounts of data, even because this would be inefficient and very expensive. Moreover, the public nature of the blockchain negates the possibility to use it for storing personal or sensitive information. For this reason, blockchain is usually adopted to notarize the existence of a document and preserve its integrity without revealing its content.
The corresponding data are then securely stored in an off-chain repository or in distributed file systems such as IPFS (InterPlanetary File System).

Data certification is essential to ensure authenticity and integrity of information, increasing trust, and reducing the risks of fraud or manipulation. Moreover, it ensures data integrity, making any changes to the original content easily detectable.
However, when data are stored outside the blockchain and need to be certified in the blockchain, the problem of keeping the certification synchronized with the data themselves arises, especially when an almost continuous flow of data needs to be certified, as in manufacturing process traceability applications.
In addition, the cost of certification must be taken into account in applications of this kind, because every transaction made on blockchain has a cost, and making a very large number of transactions could result in costs that are too high for this kind of applications.
It therefore becomes necessary to design a system that interposes itself between the data source and the blockchain, and allows for an efficient and cost-effective data certification process, without sacrificing the decentralization and certification granularity features that blockchain can offer.

\subsection{Our contribution}

In this paper, we present a blockchain-based data certification model and propose and compare some possible strategies for generating certification data at regular intervals. We analyze and evaluate different approaches in order to identify the most efficient strategy. 
Specifically, we examine two approaches: one where a blockchain transaction is created for each piece of data to include its hash, and another that leverages Merkle trees to organize document hashes, thereby reducing the number of transactions required. We also propose a cost analysis that is useful to design the most suitable system for data certification according to the final user requirements.  For the second mentioned approach, we explore the implementation of Merkle tree generation and Merkle proof extraction processes, addressing practical challenges such as handling transaction refusals during the upload of new data for certification. Transaction failures, commonly arising from fluctuating gas prices and network congestion, can significantly increase costs and introduce inefficiencies, with recent analyses indicating relatively high failure rates during periods of high congestion\footnote{See transaction failure rate analysis at: \url{https://dune.com/queries/2839305/4741578}}. 
  
  Finally, we explore a use case focused on certifying data related to event attendance and attraction visits. Finally, we present numerical results, focusing on the average execution time of the processes described above.

\subsection{Related works}

The characteristics of blockchain technology, such as immutability, integrity and time-stamping, well meet the requirements of a traceability system. For this reason, many different approaches for blockchain-based traceability are proposed in the literature.
For example, several works are focused on the use of blockchain technology for food traceability \cite{mendi2022, ibm, caro2018} and for luxury items supply chain management \cite{CHOI201917}. Moreover, different approaches for blockchain-based academic certificate management have been proposed \cite{dias2018, rustemi2023}, including some general platforms for data certification \cite{Blockcerts}. The authors in \cite{nyaletey2019} explored how to integrate blockchain with a distributed file system, aiming to provide a system capable of combining the security of the former with the efficiency of the latter.
The system proposed in \cite{Costa2022} relies on an off-chain database for data storage and on a permissioned blockchain for data certification of manufacturing data. Although the blockchain properties are maintained, users requesting a certification verification still have to trust the nodes of the network.
Regarding implementation aspects, in particular Merkle tree traversal, relevant alternatives are proposed in \cite{Jakobsson, Szydlo2004, BERMAN}.

In this paper, we consider a general approach for data traceability using blockchain. In fact, the system we consider can be efficiently applied to different amount and different kinds of data to be certified, as well as in several applications. The approaches in the literature are either specifically designed for an application or, if more general, are often complex to use or proprietary. Moreover, the system we consider uses a public blockchain, while most blockchain-based traceability systems in the literature use private or permissioned blockchains.

\subsection{Paper outline}

The paper is organized as follows. In Section \ref{sec:sysmode}, we outline the system model and evaluate the performance of two data certification approaches in terms of generation, storage, transactions, and verification cost. In Section \ref{sec:merkle}, we focus on the Merkle tree-based approach and discuss implementation aspects, also considering a use case concerning Proof of Attendance. Section \ref{sec:numres} contains numerical results, and Section \ref{sec:concl} concludes the paper.

\section{System model and certification approaches}\label{sec:sysmode}

\begin{figure}[ht]
    \centering
    \includegraphics[width=1\linewidth]{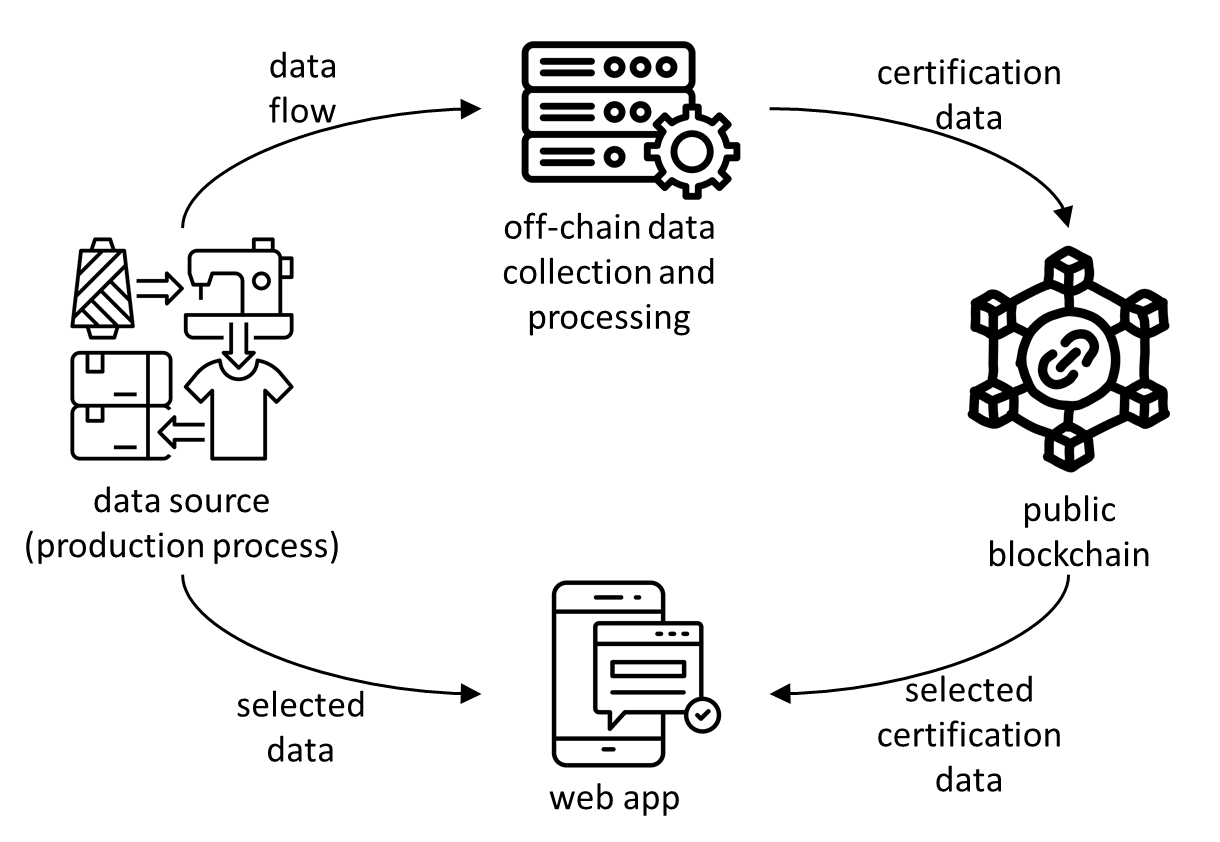}
    \caption{Blockchain-based data certification system model.}
    \label{fig:systemmodel}
\end{figure}

We consider a system for blockchain-based data certification modeled as in Fig. \ref{fig:systemmodel}.
Its main components are:
\begin{itemize}
    \item One data source, typically consisting of some production process generating traceability data in an essentially continuous manner.
    \item An off-chain data collection and processing facility, which stores the data produced in off-chain storage and processes them to obtain traceability data that are notarized on the blockchain at regular intervals.
    \item A public blockchain, like Ethereum, receiving transactions including certification data that are generated by the off-chain data collection and processing facility at regular intervals.
    \item A web app allowing users to verify the certification of some selected data, for example related to a product they purchased, directly querying the data source and the blockchain, without intermediaries.
\end{itemize}

We remark that this approach can be used for different types of data to be certified (e.g., supply chain, proof of attendance, etc.).

In this paper we delve into one component of such a system, namely that related to the best strategy for generating certification data at regular intervals.
For this purpose, let us consider two different approaches for data certification. In the first approach, every transaction contains the information to be certified, while in the second approach we aggregate more data in a single transaction, through the use of Merkle trees.

\subsection{First approach: Single data certification}

The first of the two schemes we consider is the simplest and most common one. Basically, we send a blockchain transaction that contains the hash of each piece of data that needs to be certified.
The hash of the single piece of data is included in the ``data" field of the transaction, allowing the information to be immutably stored into the blockchain. Then, the certified data are enriched with some parameters related to the blockchain transaction, such as the transaction ID, that are needed to verify data integrity. 

In fact, the verification phase constitutes a crucial aspect of a certification protocol. 
In this case, when a user wishes to verify a specific piece of data, the system takes that information together with transaction data as input, computes the hash digest, locates the transaction on-chain and compares the digest stored inside the data field of the transaction with the one computed locally. 
If the two values match, the certification is confirmed as valid, attesting data integrity. Conversely, if they do not match, the integrity verification fails, indicating a possible alteration of the information.

\subsection{Second approach: Multiple data certification}

In the second approach,  Merkle trees are used as a solution for organizing and compress the document digests.
The primary goal of this approach is to limit the number of blockchain transactions to be generated, making the certification process more efficient and cost-effective.
Compared to the previous approach, which requires a separate transaction for each piece of data to be certified, using a Merkle tree allows multiple data to be certified in a single transaction, which simply contains the root of the Merkle tree, thus saving considerable cost and processing time.
This approach requires to collect data coming from the source and organize them into Merkle trees, whose numerosity (i.e., the number of leaves) is a design choice that takes into account various factors, with the final aim of minimizing costs.
Having done so, each certified information is provided with a Merkle proof, which is a string that represents all the information participating in the Merkle tree, other than the piece of data itself. The Merkle proof allows the integrity of the information to be verified from the information itself and the Merkle tree root.
To avoid the unnecessary occupation of storage space due to storing the Merkle proof for every data leaf, we can store the entire Merkle tree structure and compute Merkle proofs on the fly when some data item needs to be associated with its Merkle proof.
Saving the tree also allows the nodes that are part of the proof to be quickly identified, speeding up the local root calculation during the verification phase.

The verification phase, indeed, is more complex than for the previous approach. In fact, verification requires local computation of the Merkle root, starting with the document whose certification is to be verified and the corresponding proof.
The proof is obtained after locating the correct tree, the position of the considered data, and all the nodes of the tree needed to re-calculate the root.
When the user has the proof and the data needed to locate the blockchain transaction (which contains the original on-chain Merkle root), the actual verification function begins, following these steps:

\begin{enumerate}
    \item Compute the hash digest of the data.
    \item Extract the original Merkle root from the blockchain transaction associated to that piece of data.
    \item Locally compute the root through the Merkle proof and the digest found in step 1$)$.
    \item Compare the values obtained in steps 2$)$ and 3$)$; if they match, the verification is successful, otherwise fails.
\end{enumerate}

\subsection{Comparison of the proposed approaches}

In the above sections we have discussed two approaches for data certification: one based on individual transactions for each data entry and the other based on data organization in Merkle trees. 
Although both methods have the same purpose, there are significant differences in terms of cost and performance.
The parameters considered for the performance evaluation are:
\begin{itemize}
    \item \textit{Generation cost}. Given $N$ data items to be certified, the first approach requires to compute $N$ hash digests and to include them in $N$ transactions. In the second approach, instead, the Merkle tree needs to be built, meaning that the number $S$ of digests to be computed is given by ${S} = (2N - 1)$. 
    Therefore, the generation cost $C_\mathrm{generation}$ in the first case is given by $C_\mathrm{generation} = N \cdot C_\mathrm{hash}$, where $C_\mathrm{hash}$ is the cost of computing one hash digest, while in the second case $C_\mathrm{generation} = ({2N - 1}) \cdot C_\mathrm{hash}$. 
    \item \textit{Storage cost}. In the first approach, no data needs to be stored. When a user wants to verify some information, they provide as input the information itself, containing the related transaction data. This information is then hashed, and the result is compared with the data contained in the blockchain transaction. As for the second approach, instead, we need to provide a Merkle proof (and possibly store the Merkle tree to improve performance). 
    Therefore, the storage cost $C_\mathrm{storage}$ is given by $C_\mathrm{storage} = S \cdot d$, where $d$ is the average dimension of the leaf node. 
    \item \textit{Transactions cost}. It is straightforward to note that the first approach has higher monetary cost that the second approach. Indeed, in the first case, we send $N$ transactions for $N$ data items, while in the second case we only need one transaction for the overall $N$ data items. Therefore, the transactions cost will be $C_\mathrm{transaction} = N \cdot p$, where $p$ is the average cost for sending a transaction in a public blockchain. The transaction cost of the second approach will instead be just $C_\mathrm{transaction} = p$. 
    We would like to underline that we consider a public blockchain because it provides higher resilience and decentralization with respect to a private or permissioned blockchain.
   \item \textit{Verification cost.} The cost of verifying a piece of data previously certified is different for the two approaches. In the first case, the verification cost is simply the cost of computing one hash digest, that is, $C_\mathrm{verification} = C_\mathrm{hash}$. As for the second approach, the Merkle proof needs to be computed. The complexity of this operation is in the order of $O(\log_{2}(N)) $; therefore, in this case, we assume $C_\mathrm{verification} = \log_{2}(N) \cdot C_\mathrm{hash}$.
\end{itemize}

Summarizing, the total cost is given by: 
$$ C_{\mathrm{tot}} = C_{\mathrm{generation}} + C_{\mathrm{storage}} + C_{\mathrm{transaction}} + C_{\mathrm{verification}}.$$
For the first approach, i.e., the single data certification, the total cost is hence given by: 
$$ C_\mathrm{tot} = N \cdot C_{\mathrm{hash}} + N \cdot p + C_{\mathrm{hash}}.$$
Instead, the total cost for the multiple data certification, requiring the Merkle tree calculation, would be:
$$ C_\mathrm{tot} = (2N - 1) \cdot C_{\mathrm{hash}} + S \cdot d + p + \log_2(N) \cdot C_{\mathrm{hash}}.$$

Looking at these overall costs, we can conclude that there is not an a priori ideal configuration, but the most suitable approach should be identified on the basis of the specific application and its requirements.
However, one aspect that must also be considered is that, in some applications, the monetary cost of transactions might be prioritized over the computational cost, and in that case the second approach is certainly preferable, as it reduces the number of transactions made on the blockchain by a factor of $N$.
In particular, in traceability applications, we usually have an almost continuous flow of data needing to be certified. Therefore, the approach that uses a Merkle tree appears to be preferable, since it minimizes the transaction costs.
For this reason, in the following sections, we provide an efficient implementation of the data certification approach based on Merkle trees. We also analyze the performance of the proposed approach in constructing a Merkle tree with $N$ data entries to be certified, as well as its performance in verifying previously certified information.

\section{Merkle Tree Generation and Merkle proof extraction}\label{sec:merkle}

Let us consider a Merkle tree implementation based on recursion and nodes indexing to keep the structure simple, easily understandable, and cost efficient. In particular, we focus on the analysis of the two most common operations applied on such a structure, which are the Merkle tree generation and the Merkle proof extraction. By leveraging the use of indexing, we guarantee low traversal costs from the root of the tree to the interested leaf.  Each node in a Merkle tree with $N$ leaves and $M$ nodes, denoted as $n_i$, $i \in \{1,\ldots,M\}$,  is defined by: 
\begin{itemize}
    \item $\mathrm{idx}_i$ such that $1\leq \mathrm{idx}_i \leq M$, the node index.
    \item $v_i$, the hash computed on the concatenation of the hashes of $n_i$ children. If $n_i$ is a leaf node, $v_i$ is the hash of the data block.
    \item $n^l_{i}$, the left child (null if $n_i$ is a leaf).
    \item $n^r_{i}$, the right child (null if $n_i$ is a leaf).
\end{itemize}

\subsection{Nodes indexing when $N$ is a power of $2$\label{subsec:pow2}}
\begin{figure}[th!]
    \centering
    \begin{adjustbox}{valign=t}
    \begin{forest}
      [8
        [4 [2[1][3]] [6[5][7]] ]
        [12   [10 [9][11]] [14 [13][15]] ]
      ]
    \end{forest}
    \end{adjustbox}\qquad
    \begin{adjustbox}{valign=t}
    \begin{forest}
    for tree={
        edge={draw=none} 
      }
    [\textit{(h=3)}[\textit{(h=2)}[\textit{(h=1)}[\textit{(h=0)}]]]]
    \end{forest}
    \end{adjustbox}
\caption{Example of indexing of a Merkle Tree with 8 leaves.}
\label{fig:indexed_mktree}
\end{figure}
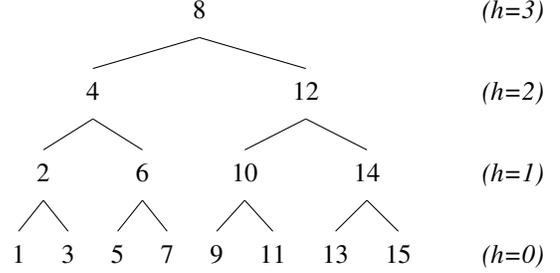

The tree indexing when $N$ is a power of $2$ is performed using a bottom-up approach according to the following rules:
\begin{itemize}
    \item The indexes of the leaves are defined as incrementing odd numbers, starting from 1 on the leftmost leaf.
    \item The index of each parent node $p$ is computed as: 
    \begin{equation}\label{index_eq}
      \mathrm{idx}_p = \mathrm{idx}_l + 2^{h},  
    \end{equation}
    where $\mathrm{idx}_l$ represents the index of the direct left child and $h$ corresponds to the height of the node children.
    For example, the index $\mathrm{idx}_p$ of the direct parent of each pair of leaves is computed as:
    \[\mathrm{idx}_p = \mathrm{idx}_l + 2^{h} = \mathrm{idx}_l + 2^{0} = \mathrm{idx}_l+1.\]
\end{itemize}
An example of indexed Merkle tree is shown in Fig. \ref{fig:indexed_mktree}, where, for the sake of conciseness, only the entry indexes are reported.

If $N$ is not a power of $2$, a modification of this indexing procedure must be performed, as discussed next.

\subsection{Merkle Tree generation procedure}

Given a list of data items and a hash function $H$, hashing is applied to each element of the list to obtain the corresponding hashed data list. Each element of the hashed data list will be a leaf node of the Merkle tree.
In general, for every pair of consecutive nodes $(n_i, n_{i+1})$ with hash values $v_i$ and $v_{i+1}$, a parent node is obtained by computing the node hash value as $H(v_i || v_{i+1})$, where $||$ denotes concatenation, and the index through  (\ref{index_eq}). Throughout the process, we keep two lists which are refreshed at every tree climb: a \textit{values list} $L_v$ and a \textit{nodes list} $L_n$. $L_n$ contains the information regarding the nodes created at the current layer. If we are not at the root layer, these nodes will be included as left and right children of their parent nodes. $L_v$, instead, contains the hash values $v_i$ of the nodes created at the current layer.  If we are not at root level, these hashes are used to compute the hash value of their parent nodes.

\begin{figure}[ht]
    \centering
    \begin{adjustbox}{valign=t}
    \begin{forest}
      [8
        [4 [2[1][3]] [6[5][7]] ]
        [12   [10 [9][9]] [10 ] ]
      ]
    \end{forest}
    \end{adjustbox}\qquad
    \begin{adjustbox}{valign=t}
    \begin{forest}
    for tree={
        edge={draw=none}
      }
    [\textit{(h=3)}[\textit{(h=2)}[\textit{(h=1)}[\textit{(h=0)}]]]]
    \end{forest}
    \end{adjustbox}
\caption{Example of indexing of a Merkle Tree with 5 leaves.}
\label{fig:unbalanced_mktree}
\end{figure}
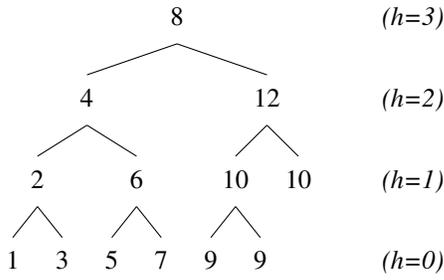

It can happen that the number of leaves is not a power of $2$. This implies that at least a (non-root) layer contains an odd number of nodes. In our implementation, we need to bypass such a condition. So, before climbing the tree, if the length of $L_n$ is odd and greater than $1$, the last node of $L_n$ is replicated, and its left and right children set as zero, as shown in Fig. \ref{fig:unbalanced_mktree}.

When all the nodes of the current step have been processed, we move on  to the next layer. Having $L_v$ and $L_n$ filled, the procedure iterates through such lists as follows. Let us denote as $n^l_{p}$ and $n^r_{p}$ the left and right child of a given parent node $n_p$, respectively. Then, 
\begin{enumerate}
    \item for every two consecutive elements of $L_n$, say $n_i$ and $n_{i+1}$, it holds that $n_i = n^l_{p}$ and $n_{i+1} = n^r_{p}$ for a new parent node $n_p$;
    \item $n_p$ hash value is computed as $H(L_{v}[i] \:||\: L_{v}[i+1])$;
    \item $n_p$ index is computed with  (\ref{index_eq}).
\end{enumerate}
When the lists $L_v$ and $L_n$ have been completely explored, they are cleared. Such procedure goes on until $L_v$ and $L_n$ are populated with only one element, which is the Merkle root. The pseudo-code is shown in Algorithm \ref{alg:merkcos}, which includes the indexing procedure described in Section \ref{subsec:pow2} as a special case. The Merkle tree generation has a time complexity in the order of $O(N)$.

\begin{algorithm}
\caption{Merkle Tree Construction}
\label{alg:merkcos}
\begin{algorithmic}[1]
\renewcommand{\algorithmicrequire}{\textbf{Input:}}
\renewcommand{\algorithmicensure}{\textbf{Output:}}
\Require List of data items $D = \{d_1,  \ldots, d_n\}$, hash function $H$
\Ensure A properly constructed Merkle tree with root node

\State \textbf{Initialize:}
\State $L_v \gets \emptyset$ \Comment{Values list}
\State $L_n \gets \emptyset$ \Comment{Nodes list}

\State \textbf{Hash input data:}
\For{$i \gets 1$ to $n$}
    \State $v_i \gets H(d_i)$ \Comment{Hash each data item}
    \State Create leaf node $n_i$ with index $2i-1$ and value $v_i$
    \State $L_v$.append($v_i$)
    \State $L_n$.append($n_i$)
\EndFor

\State \textbf{Build tree bottom-up:}
\State $h \gets 0$ \Comment{Current height}
\While{$|L_n| > 1$} \Comment{Continue until reaching the root}
    \State $L_v^{new} \gets \emptyset$ \Comment{New values list for next level}
    \State $L_n^{new} \gets \emptyset$ \Comment{New nodes list for next level}
    
    \If{$|L_n|$ is odd \textbf{and} $|L_n| > 1$}
        \State Duplicate last node in $L_n$ 
        \State Duplicate last value in $L_v$
    \EndIf
    
    \For{$i \gets 0$ to $|L_n|-1$ \textbf{by} 2} 
        \State $n_i \gets L_n[i]$ \Comment{Left child}
        \State $n_{i+1} \gets L_n[i+1]$ \Comment{Right child}
        \State $v_i \gets L_v[i]$ \Comment{Left child hash value}
        \State $v_{i+1} \gets L_v[i+1]$ \Comment{Right child hash value}
        
        \State $v_p \gets H(v_i \parallel v_{i+1})$ \Comment{Compute parent hash value}
        \State $idx_p \gets n_i.index + 2^h$ \Comment{Calculate parent index using (\ref{index_eq})}
        
        \State Create parent node $n_p$ with index $idx_p$, valued $v_p$
        \State Set $n_i$ as left child of $n_p$
        \State Set $n_{i+1}$ as right child of $n_p$
        
        \State $L_v^{new}$.append($v_p$)
        \State $L_n^{new}$.append($n_p$)
    \EndFor
    
    \State $L_v \gets L_v^{new}$ \Comment{Update lists for next level}
    \State $L_n \gets L_n^{new}$
    \State $h \gets h + 1$
\EndWhile

\State \Return Indexed Merkle tree, $L_n[0]$
\end{algorithmic}
\end{algorithm}

\subsection{Proof extraction}
\begin{figure}[ht]
    \centering
    \begin{adjustbox}{valign=t}
    \begin{forest}
      [8
        [\textcolor{red}{4} [2[1][3]] [6[5][7]] ]
        [12   [\textcolor{red}{10} [9][11]] [14 [13,circle,draw, inner sep=1pt][\textcolor{red}{15}]] ]
      ]
    \end{forest}
    \end{adjustbox}\qquad
    \begin{adjustbox}{valign=t}
    \begin{forest}
    for tree={
        edge={draw=none} 
      }
    [\textit{(h=3)}[\textit{(h=2)}[\textit{(h=1)}[\textit{(h=0)}]]]]
    \end{forest}
    \end{adjustbox}
\caption{Example of proof extraction of leaf with index 13. The indexes in red represent the nodes whose hash digests are collected to compose the Merkle proof.}
\label{fig:mk_proof}
\end{figure}
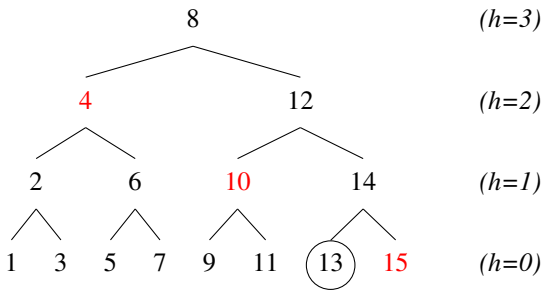

Having built a Merkle tree and knowing the index of the desired leaf, the proof can be extracted by traversing the tree from the root to the leaf following the nodes indexes as follows. We define as $\mathrm{idx}_i$ the index of the leaf and with $\mathrm{idx}_n$ the index of the node that is being traversed. Then,
\begin{itemize}
    \item if $\mathrm{idx}_i > \mathrm{idx}_n$, store the hash value of the left child node and move on the right sub-tree;
    \item if $\mathrm{idx}_i < \mathrm{idx}_n$, store the hash value of the right child node and move on the left sub-tree;
    \item else, the leaf has been found.
\end{itemize}
Such an operation has a time complexity in the order of $O(h) = O(\log_{2}(N))$, where $h$ represents the tree height and $N$ represents the number of leaves.
An example of proof extraction through nodes traversing with indexes is shown in Fig \ref{fig:mk_proof}.

\subsection{Multi-level Indexed Merkle Tree}

Basically, the system considers that a general transaction workflow would be built over a waiting queue mechanism in which, for a defined time, clients can submit data to be certified. After a trigger is activated, which can be automatic (for example, time-based) or manual (certification request), the Merkle tree is built and saved off-chain, then the Merkle root is inserted in a data transaction which is submitted to the network. Using Ethereum, the client requesting the certification must define a gas price that they are willing to pay. If the gas price is not enough, the transaction gets refused by the network and the client must apply again for a new certification request after having tuned the gas price. In the meantime, it could happen that new data to be certified have been collected, and such data are not included in the Merkle tree related to the refused transaction. 
The simplest solution would be to generate a new Merkle tree and a new transaction, but since the previous one has not yet been written onto the blockchain, a more efficient solution can be adopted. In particular, it is possible to incorporate certification of the new data into the previous transaction, which can be updated before being resubmitted.

We propose a solution that addresses this issue by building a Merkle tree that can be extended without requiring a rearrangement in leaves positioning. This way, even if the network rejects the data transaction several times, it is possible to add new data without having to build another Merkle tree or assign updated indexes to previously uploaded data. Although leaf rearrangement is a simple solution, it cannot be used when the leaf position of a data item needs to be known at the time of data storage, for example to extract the Merkle proof later without having to explore all the leaves. By adopting a multi-level approach, such a requirement is fulfilled. In fact, the predefined leaf index, assigned when generating the Merkle tree for a failing transaction, is still valid when generating a new multi-level indexed Merkle tree, as explained next.

\begin{figure}[h!]
    \centering
    \begin{adjustbox}{valign=t}
    \begin{forest}
      [4
        [2, s sep=7mm
        [\textit{1,4},tikz={\node [draw,red,inner sep=0,fit to=tree]{};}
            [\textit{1,2}[\textit{1,1}][\textit{1,3}]] [\textit{1,6}[\textit{1,5}][\textit{1,7}]] ]
        [\textit{3,2},tikz={\node [draw,blue,inner sep=0,fit to=tree]{};}
            [\textit{3,1}] 
            [\textit{3,3}]
        ]]
        [6, s sep=7mm
        [\textit{5,2},tikz={\node [draw,green,inner sep=0,fit to=tree]{};}[\textit{5,1}][\textit{5,3}]] 
        [5]
      ]]
    \end{forest}
    \end{adjustbox}\qquad
\caption{Example of a multi-level Merkle tree obtained before attempting to forward the transaction for the third time. Each colored box represents a different indexed Merkle tree.}
\label{fig:multi-lev-mktree}
\end{figure}
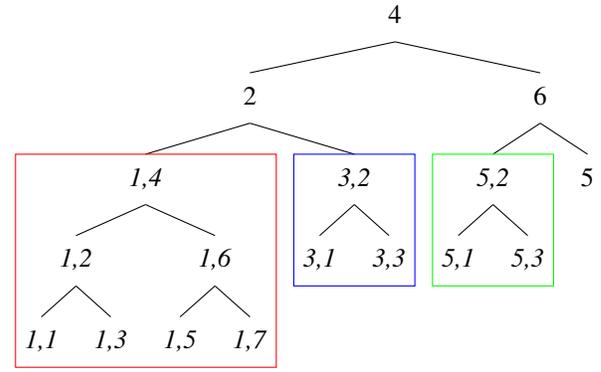

We propose a multi-level indexed Merkle tree that consists of a combination of single indexed Merkle trees, built on different levels (see Fig. \ref{fig:multi-lev-mktree} for an example). Let us assume that the Merkle tree containing the highest layer is labeled as level-$0$ Merkle tree. We say that a Merkle tree lives in the $l$-th level if index representing its root is a vector with $l+1$ entries. The number of levels is predetermined and depends on the considered application. The last level is the one dealing with transaction failures and, in practice, the number of subtrees it will contain by the end of the certification process cannot be known.

For example, let us consider a $2$-level indexed Merkle tree, containing $m$ subtrees at level $1$. The level-$1$ subtrees are indexed first, as follows. Let $n_w$ be a node of the $z$-th subtree, $0\leq z\leq m-1$, then the index $\mathrm{idx}_w$ of such node is $\mathrm{idx}_w = [2z+1,j]$, where $j$ is found according to \eqref{index_eq}. Once a subtree is fully constructed, it is appended to the level-$0$ Merkle tree, with its root as a leaf node. Then, the indexing of the  level-$0$ Merkle tree is again carried out following \eqref{index_eq},  with the right entry of the indexes of the leaf nodes being ignored during this step.

This reasoning easily generalizes to $k$-level Merkle trees. Moreover, with this approach, indexes are resistant to transaction failures. In fact, the leaf index assigned to the data already has a complete path traversal considering all the tree levels. Let us assume we adopt a $2$-level tree where level $1$ manages transaction failures and level $0$ is used to obtain the final Merkle root. If there were $m-1$ transaction failures, the whole multi-level indexed tree will contain $m$ subtrees at level-$1$. Since the subtrees are incrementally appended on an ``as needed'' basis, it is possible to obtain the definitive leaf index regardless of the final value of $m$.

\subsection{Use case: Proof of Attendance}


As stated in the Introduction, an interesting application for traceability using blockchain is related to the Proof of Attendance, i.e., certifying users’ participation in various kinds of events. 
For this reason, the infrastructure described in the previous sections has been applied to a use case regarding the blockchain-based certification of participation in cultural events and touristic initiatives. In both cases, when attending an event or visiting an attraction, participants are asked to prove their attendance by scanning some QR code. Since events have a limited time for attending, the certification on blockchain is done only after the event itself is ended by generating a single Merkle tree, where each leaf corresponds to a participant. For attractions, instead, since there is no time limit, the attractions manager can decide when to apply for certification. The problem related to the transaction failure, thoroughly described in the previous section, can occur in the latter case. Let us suppose that a set of attendants have registered their presence and are waiting for their attendance to be certified, the attractions manager applies for a transaction on blockchain, but the transaction fails. In the meantime, other attendants might have scanned the QR code. As a result, if no action is undertaken to modify the Merkle tree approach, it would be necessary to apply for two transactions and spend a double price: one for the failed transaction, which needs to be forwarded again, and one for the new one. Generally, an attractions manager might have multiple attractions with different QR codes. Attendants of each event would be managed through different Merkle trees. When paying for attendance certification, the attractions owner would need to pay a price 
\[P = p \cdot \sum_i(1 + F_i)\]
with $p$ being the price of a data transaction, $i$ being an index for the attractions, and $F_i$ being the number of transactions failed for the $i$-th attraction.
By leveraging the multi-level indexed Merkle tree it is possible to minimize $P$ by generating a single Merkle tree as follows:
\begin{itemize}
    \item level $0$ is the top-level Merkle tree, which contains the final Merkle root to be forwarded to the blockchain;
    \item level $1$ contains the sub-trees $a_i$ representing the attractions;
    \item level $2$ contains the sub-trees $t_{i,j}$, representing failed transactions (with index $j$) for the $i$-th attraction.
\end{itemize}

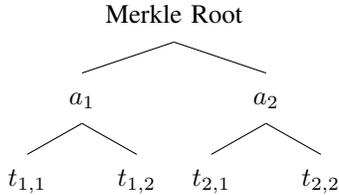
\begin{figure}[h!]
    \centering
    \begin{adjustbox}{valign=t}
    \begin{forest}
      [Merkle Root
        [$a_1$, s sep=7mm
        [$t_{1,1}$]
        [$t_{1,2}$
        ]]
        [$a_2$, s sep=7mm
        [$t_{2,1}$] 
        [$t_{2,2}$] 
      ]]
    \end{forest}
    \end{adjustbox}\qquad
\caption{Example of a multi-level Merkle tree used in the Proof of Attendance use case. }
\label{fig:poa}
\end{figure}

As a result it holds that \(P = p\), independent of the number of failures. In Fig. \ref{fig:poa}, an example of tree for the management of the attendance for two attractions is proposed. In particular, in the figure, the tree is obtained before attempting to forward the transaction for the second time for certificating attendance at two events. The sub-trees $t_{1,1}$ and $t_{2,1}$ contain attendances related to a previously failed transaction. If $a_2$ did not have any attendants when the failed transaction was requested, but some attendants registered before the second attempt, $t_{2,1}$ will contain such attendants and $t_{2,2}$ would only contain the Merkle Root of $t_{2,1}$ for balancing purposes.

\section{Numerical results}\label{sec:numres}

\begin{figure}[ht]
    \centering
    \includegraphics[width=1\linewidth]{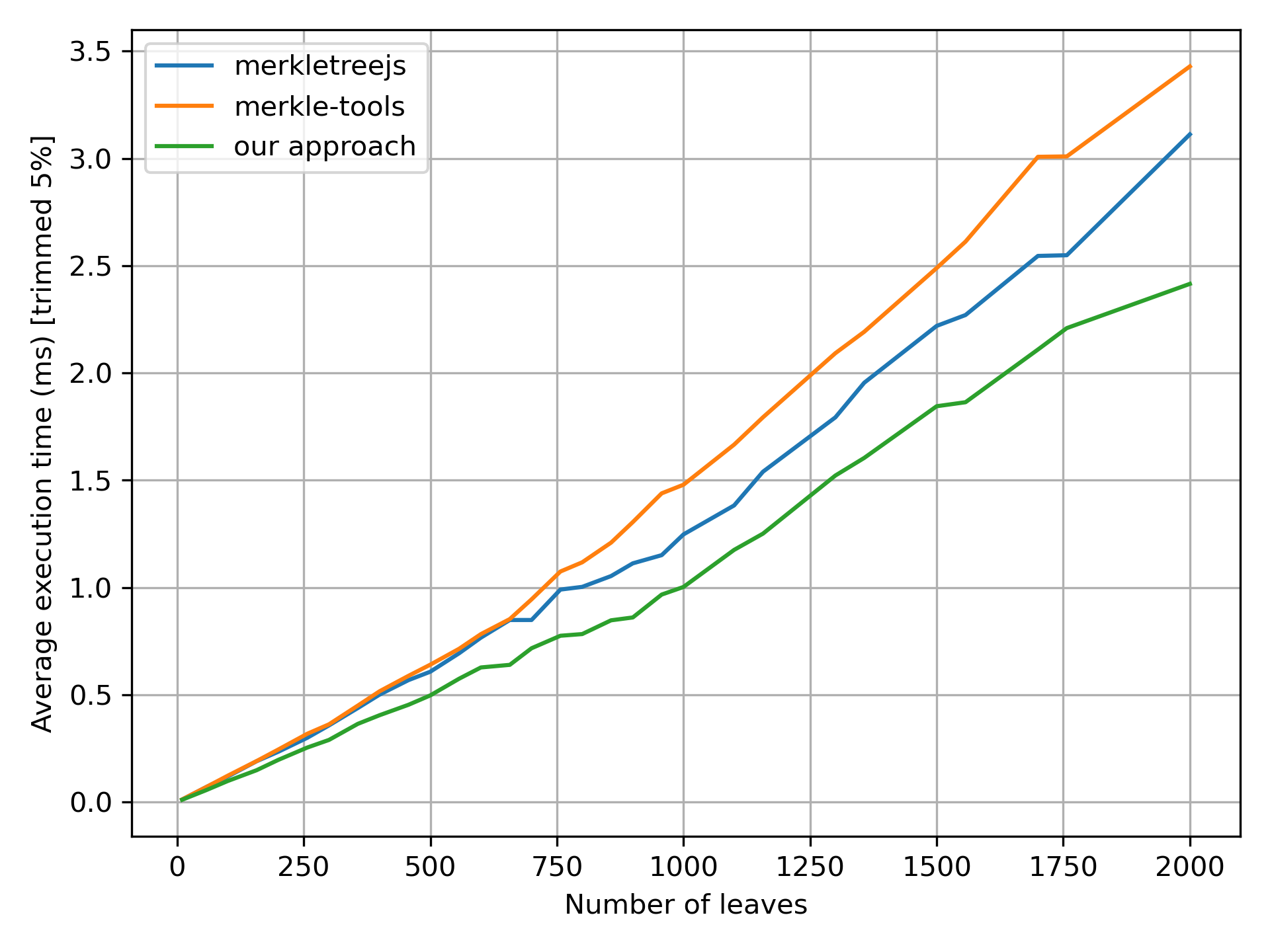}
    \caption{Merkle tree generation performance}
    \label{fig:ext1}
\end{figure}

To evaluate the performance of the proposed indexed Merkle tree, we compare our approach with two public libraries: \textit{merkletreejs} \cite{mktreejs} and \textit{merkle-tools} \cite{mktools}. In particular, we compare the average times required to execute the Merkle tree generation and the Merkle proof extraction. All  tests were carried out on an Apple Macbook Pro 2021 with a Apple M1 Pro chip and 16 GB of RAM. 
First, to limit the variability caused by the random leaves data generation, we generate leaves data only once and then we start measuring performance on the aforementioned operations. We assess the time complexity in relation to the number of leaves and we do it by applying the same function for $250\,000$ iterations. In our tests, we consider the following leaf sizes: 
 \{$10$, $57$, $100$, $157$, $200$, \dots, $900$, $957$, $1000$, $1100$, $1157$, $1300$, $1357$, $1500$, $1557$, $1700$, $1757$, $2000$\}. On the results obtained for each size, we apply a $5\%$ trimmed mean on measured times. For both operations tested, we report a figure representing the overall results as the number of leaves increases, and Table \ref{tab:build}, analyzing execution times, together with a percentage increase comparing the libraries with our approach..
The results reported in Fig. \ref{fig:ext1} show that all the approaches follow a trend strictly related to the number of leaves. Compared to existing libraries, our approach requires less time to build a tree, and such a difference becomes more evident as the number of leaves increases. Our approach shows higher efficiency, with \textit{merkletreejs} being slower by a quantity ranging between $19.32\%$ and $28.86\%$, and \textit{merkle-tools} having greater delays ranging between $24.68\%$ and $47.55\%$ (see Table \ref{tab:build}).

\begin{table}[ht]
\caption{Merkle tree generation performance}
    \centering
    \begin{tabular}{c c c c}
          &  \multicolumn{3}{c}{\textbf{Average execution time (ms)}}\\
          
         $N$ & merkletreejs \cite{mktreejs} & merkle-tools \cite{mktools} & our approach \\
         \cmidrule{1-4}
         100 &  0.1182 (+21,23\%) &  0.1222 (+25,35\%) & 0.0975\\
         200 & 0.234 (+19,32\%) & 0.2445 (+24,68\%) & 0.1961\\
         500 & 0.6071 (+22,25\%) & 0.6399 (+28,85\%) & 0.4966\\
         1000 & 1.2471 (+24,44\%) & 1.4788 (+47,55\%) & 1.0022\\
         1500 & 2.2192 (+20,3\%) & 2.489 (+34,93\%) & 1.8447\\
         2000 & 3.1123 (+28,86\%) & 3.43 (+42,02\%) & 2.4152\\
         
    \end{tabular}
    \label{tab:build}
\end{table}

\begin{figure}[ht]
    \centering
    \includegraphics[width=1\linewidth]{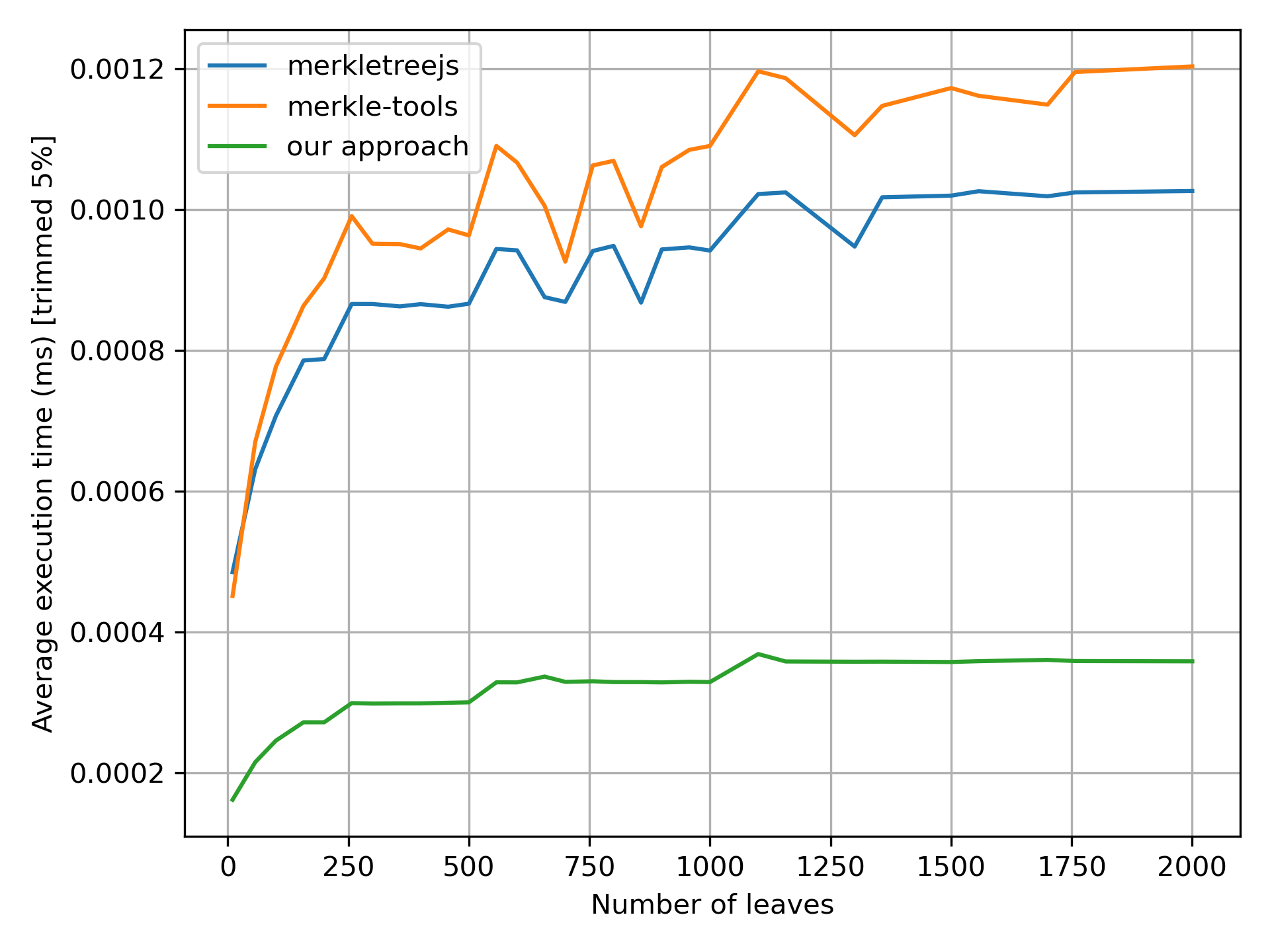}
    \caption{Proof extraction comparison}
    \label{fig:ext2}
\end{figure}

Fig. \ref{fig:ext2} shows the results of the Merkle proof extraction. Owing to the index-based exploration of the tree, it is possible to maintain a lower execution time, that scales rather slowly with the number of leaves. Differences with \textit{merkletreejs} and \textit{merkle-tools} are more noticeable with delays in time ranging from $177.78\%$ to $200\%$ for the first one and from $181.25\%$ to $233.33\%$ for the second one (see Table \ref{tab:proof}). 

\begin{table}[ht]
    \centering
    \caption{Merkle proof extraction performance}
    \begin{tabular}{c c c c}
          
          &  \multicolumn{3}{c}{\textbf{Average execution time (ms)}}\\
          
         $N$ & merkletreejs \cite{mktreejs} & merkle-tools \cite{mktools} & our approach \\
         \cmidrule{1-4}
         100 & 0.00048 (+200\%) & 0.00045 (+181,25\%) & 0.00016\\
         200 & 0.00079 (+192,6\%) & 0.0009 (+233,33\%) & 0.00027\\
         500 & 0.00087 (+190\%) & 0.00096 (+220\%) & 0.0003\\
         1000 & 0.00094 (+184,85\%) & 0.0011 (+233,33\%) & 0.00033\\
         1500 & 0.00102 (+183,33\%) & 0.00117 (+225\%) & 0.00036\\
         2000 & 0.001 (+177,78\%) & 0.0012 (+233,33\%) & 0.00036\\
         
    \end{tabular}
    \label{tab:proof}
\end{table}

These results show that, from a data certification perspective, where an agile Merkle proof extraction needs to be performed, the use of the indexed Merkle tree can significantly reduce the execution time.

\section{Conclusion}\label{sec:concl}

We have studied some issues related to blockchain-based data certification, with particular reference to traceability applications, where an almost continuous flow of data needs to be certified by writing certification data onto a public blockchain.
We considered and compared two possible strategies for organizing the data to be certified and computing the related certification information. The second one, in particular, makes use of Merkle trees and is particularly suitable for keeping the monetary cost of blockchain transactions low when the data to be certified are many and generated in an almost continuous manner.
We have also considered a practical use case, concerning the certification of attendance at cultural events or tourism initiatives, showing an efficient implementation of Merkle trees generation and Merkle proof extraction.
The techniques considered have also been experimentally evaluated, and our numerical results show the practical feasibility and advantages of our approach, even in the presence of a continuous flow of data to be certified.
Last but not least, we showed how the use of multi-level Merkle trees can allow managing transaction failures in an efficient way.
As a hint for possible future work, we expect to perform an in-depth security analysis of the system, possibly using automated assessment methods. Moreover, we are evaluating the use of efficient Merkle tree implementations for decentralized authentication based on fuzzy sources \cite{Alzahab2024,Santini2023}.


\bibliographystyle{IEEEtran}
\bibliography{biblio}

\end{document}